\documentclass[jcp,aip,groupedaddress,amsmath,reprint]{revtex4-1}

\usepackage{graphicx}
\usepackage{dcolumn}
\usepackage{color}
\usepackage{bm}
\usepackage{amsmath}
\usepackage{amsfonts}
\usepackage{amssymb}
 \newcommand{\be}{\begin{equation}}
 \newcommand{\ee}{\end{equation}}
\newcommand{\bea}{\begin{eqnarray}}
\newcommand{\eea}{\end{eqnarray}}
\newcommand{\ba}{\begin{eqnarray*}}
\newcommand{\ea}{\end{eqnarray*}}

\newcommand{\bx}{\mathbf{x}}

\usepackage{multirow}

\begin{document}

%%%%%%%%%%%%%%%%%%

%\widetext

%\twocolumn
%[\hsize\textwidth\columnwidth\hsize\csname
%@twocolumnfalse\endcsname
%%%%%%%%%%%%%%%%%%

\title{Accelerated nuclear quantum effects sampling with open path integrals} 

\author{Guglielmo Mazzola}
 \email[]{gmazzola@phys.ethz.ch}
 \affiliation{Theoretische Physik, ETH Zurich, 8093 Zurich, Switzerland}

\author{Matthias Troyer}

 \affiliation{Theoretische Physik, ETH Zurich, 8093 Zurich, Switzerland}
\affiliation{Quantum Architectures and Computation Group, Microsoft Research,
Redmond, WA 98052, USA}
\date{\today}

\begin{abstract}
%{\color{red} [MT: I would not start with "we study" but right away with the punchline result]} We study the autocorrelation properties of path integral Monte Carlo (PIMC) equilibrium simulations with periodic and open boundary conditions in imaginary time.
We numericaly demonstrate that, in double well models,  the autocorrelation time of open path integral Monte Carlo  simulations can be much smaller compared to standard ones using ring polymers.
We also provide an intuitive explanation based on the role of \emph{instantons} as transition states of the path integral pseudodynamics.
Therefore we propose that, in all cases when the ground state approximation to the finite temperature partition function holds, open path integral simulations can be used to accelerate the sampling  in realistic simulations aimed to explore nuclear quantum effects.
\end{abstract}

\maketitle
%]
%%%%%%%%%%%%%%%%%%
%%%%%%%%% PACS USED
%%%%  71.10.Fd Lattice fermion models (Hubbard model, etc.)
%%%%  74.20.Mn Nonconventional mechanisms (spin fluctuations, polarons and bipolarons, resonating valence bond model, anyon mechanism, marginal Fermi liquid, Luttinger liquid,
%%%%  71.10.Pm Fermions in reduced dimensions
%%%%  71.27.+a Strongly correlated electron systems; heavy fermions
%%%%%%%%%%%%%%%%%%
\section{Introduction}

Nuclear quantum effects (NQE) are of utmost importance in a broad class of compounds containing light atoms.
For example, due to the pivotal role of the hydrogen bond, the zero point motion of the protons strongly affects the description of water and related aqueous system even at room temperature\cite{ceriotti2016nuclear,carwater,ceriotti2012efficient,ceriotti2013nuclear}.
Moreover, under particular conditions, such as high pressure\cite{benoit1998tunnelling} or adsorption on surfaces\cite{meng2015direct}, also proton tunneling events\cite{richardson2016concerted} occur frequently and change the physics of the systems.
NQE are also essential for describing, even at the qualitative level, the phase diagram of high pressure hydrogen, the simplest condensed matter system\cite{bonev2004quantum,pickard_structure_2007,labet_fresh_2012,morales_nuclear_2013,mazzola_unexpectedly_2014}.
Here, the tiny free energy differences between competing crystal structures, computed with the classical nuclei approximations, implies that the inclusion of NQE reorders the energetically favourable lattices at any given pressure. The most important consequence concerns the long sought low-temperature metallization of dense hydrogen\cite{eremets_conductive_2011,howie_mixed_2012,dalladay2016evidence} which, in the solid phase, crucially depends on the lattice structure.
NQE are also important in the dense liquid phase up to temperatures of 2000 K, as they may explain residual differences between numerical simulations\cite{morales_nuclear_2013,mazzola2015distinct,sorella2016accelerated} and experiments\cite{knudson2015direct} concerning the molecular dissociation and metallization in the fluid phase. 

Path integral Monte Carlo (PIMC) and path integral molecular dynamic (PIMD) simulations are the most popular approach in realistic simulations to reproduce NQE as far as equilibrium properties are concerned.
These methods directly arise from the Feynman path integral formulation of quantum mechanics and are able to simulate exactly the quantum statistic when the distinguishable particles approximation holds, as in the above condensed matter systems examples.

To briefly introduce this technique we start from the expression for the partition function $Z$:
\begin{equation}
\label{eq:Z}
Z = \int dx \langle x | e^{-\beta H} | x \rangle
\end{equation}
where $x$ is the quantum particle coordinate (the generalization to arbitrary dimensions is straightforward), $\beta=1/k_BT$ is the inverse temperature and $H$ is the Hamiltonian of the system.
We first notice that the operator $e^{-\beta H}$ corresponds to an evolution in imaginary time $\beta$. We employ the Trotter-Suzuki approximation, which is based on the possibility to neglect to commutator between the non-commuting terms of $H = T + V$ (with $[T,V] \neq 0$), if the imaginary propagation time, $\tau$, 
 is small, i.e. $e^{-\tau (T+V)} \approx e^{-\tau T}e^{-\tau V}$.
 In typical condensed matter Hamiltonians, $T=1/2m ~ \partial^2 / \partial x^2$ is the kinetic operator, and $V(x)$ is the potential energy,  which can either be given by a empirical force fields  or by \emph{ab-initio} calculations, such as quantum Monte Carlo or density functional theory.
Splitting the imaginary time evolution into $P$ small time steps of length $\delta_\tau = \beta/P$, the path integral expression for Eq.~(\ref{eq:Z}) then becomes
 \begin{equation}
 \label{e:pi}
 Z \propto \int d x_1 d x_2 \cdots d x_P \exp{\sum_{i=1}^P S_i }~,
 \end{equation}
where $S_i = K_i + U_i$ is the \emph{action} of each step.
$K_i= (x_{i-1}-x_i)^2/(2\delta_\tau/m)$ is the kinetic part and $U_i= \delta_\tau/2(V(x_{i-1} - x_i)$, in the so-called \emph{primitive} approximation.
Notice that $x_1 = x_P$ (closed boundary conditions in imaginary time), for evaluating the trace of the density operator.   

This provides an analogy between a quantum system and a classical system with an additional dimension: Eq.~(\ref{e:pi}) is a classical configurational integral and the multidimensional object $(x_1,\cdots,x_{P-1})\equiv \bx(\tau)$ can be viewed as a \emph{ring-polymer}, whose elements are connected by springs. Each element is labeled by its position along the imaginary time axis, with $0 \leq \tau < \beta$. 
We refer to the Ref.~\onlinecite{RevModPhys.67.279} for a detailed review of path-integrals.
An essential feature of Eq.~(\ref{e:pi}) is that the integrand is always positive, and hence the distribution $\exp{\sum_{i=1}^P S_i }$ can be sampled by means of Metropolis Monte Carlo methods or Molecular Dynamics (MD) simulations. 

We note that now the computational effort is increased  by at least a factor $P$ compared to the classical nuclei approximation.
For this reason several techniques have been proposed to boost the efficiency of this approach, such as colored-noise thermostats\cite{ceriotti_nuclear_2009,ceriotti}, ring-polymer contraction approaches\cite{markland2008efficient,PhysRevE.93.043305} and multiple time-step MD\cite{kapil2016accurate}.
Here we propose a simple approach, which can be  combined with the above-mentioned techniques and can be straightforwardly applied to any existing software package for path integral simulations.
Our technique is based on simulations with open boundary conditions in imaginary time.  It is applicable  in the low temperature limit, when the thermal quantum density distribution can be safely approximated by the ground state one.

Notice that the idea of using open path integrals in the realm of realistic simulations is certainly not new.
Indeed open paths have been employed  to find the ground state -- this method was originally called \emph{path integral ground state}\cite{pigs,doi:10.1021/jp4015178,pilati2012bosonic}(PIGS) --- or to compute off-diagonal operators, such as the momentum, in helium\cite{pollock,bertaina2016one,roggero2013dynamical,cuervo2005path} or liquid water\cite{morrone2007}.
To this end they have also been used to study NQE in water in the pioneering work of Morrone and Car\cite{carwater}.
This technique is also connected with the \emph{reptation} Monte Carlo\cite{PhysRevLett.82.4745,carleo2010reptation} technique, which employs a different update scheme for the open path. The main result of our paper is that open paths could also greatly reduce autcorrelation times in path integral based simulations and thus lead to more efficient simulations

\section{Instantons in the PIMC pseudodynamics}

%Renewed efforts have been recently spent  assessing the efficiency of PIMC for the simulation of tunneling events.
Connections between  exact quantum dynamics and PIMD approaches, such as Centroid Molecular Dynamics\cite{cmd} and Ring Polymer Molecular Dynamics\cite{rpmd} have been discussed for a while \cite{voth,shorttime,hele}, and have recently gained attention in the  completely different field of adiabatic quantum optimization. There, PIMC is employed to simulate and predict the behaviour  of quantum annealing devices\cite{farhi2000quantum,Farhi20042001,johnson2011quantum}
% {\color{red}[MT: rather refer to some D-Wave papers and Farhi's papers for QA - using the original 2000 (2001?) and 2002 papers and cite our Science at NatPhys papers for applications ]}   
 which use quantum tunneling to solve combinatorial optimization problems\cite{boixo2014,ronnow2014defining}.

In particular, in Ref.~\onlinecite{isakov2015understanding}, tunneling events in a ferromagnetic ising model have been studied with PIMC.
This spin system can ne described by an effective double well model and
it has been numerically demonstrated that PIMC tunneling events occurs  with a rate $k$ which scales exactly, to leading exponential order, with the gap squared, $\Delta^2$, of the system, i.e. of the tunneling splitting energy squared.

Moreover it has been also shown that, if path integrals with open boundary conditions (OBC) in imaginary time are employed, the tunneling rate scales simply with $\Delta$, thus providing a quadratic speed-up over the standard PIMC approach. 
%This evidence  could appear surprising at first, since PIMC is a statistical method and does not solve the time-dependent Schr\"{o}dinger equation, but 
 The simple picture that has been provided in Refs.~\onlinecite{isakov2015understanding,jiang2016scaling} to understand this scaling lies into the \emph{instanton} theory of tunneling. Below we summarize  t.he results of these papers.

   \begin{figure}[t]
 \includegraphics[width=1\columnwidth]{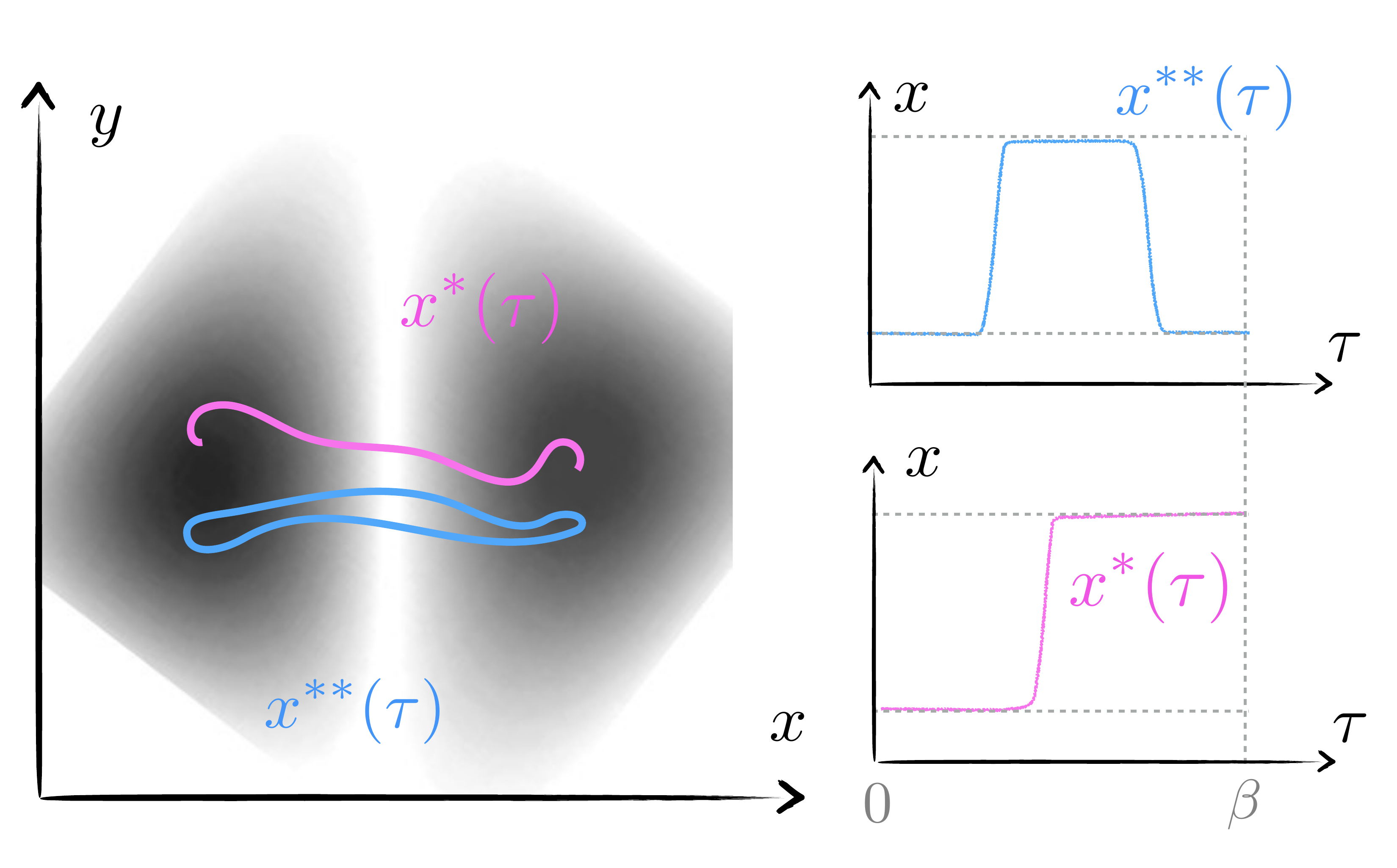}
 \caption{ (color online). \emph{Left.} Cartoon of the typical instantonic paths in configuration space, with PBC, $\bx^{**}(\tau)$ (cyan) and OBC in imaginary time, $\bx^{*}(\tau)$ (pink). These paths are transition states of the PIMC and PIGS pseudodynamics respectively (in the space of imaginary time trajectories) in double well models (sketched in the grey scale heatmap). \emph{Right.} Instantonic trajectories (projected ont ethe reaction cooordinate $x$ axis) as a function of the imaginary time $\tau$. Notice that PIMC instantons have to cross twice the barrier to fulfill the PBC constrain.
 }
 \label{fig:cart}
 \end{figure}

%Let us now return on PIMC ( or equivalently PIMD) simulations, where we samples paths $\bx(\tau,t)$ at each update along the simulation time axis $t$, and these paths are distributed according the free-energy functional $ F = \m S_\beta[\bx(\tau)]/\beta $.
%%The quantum tunneling event has therefore been recast into a thermally activated escape event, which occurs when the PIMC trajectory crosses the saddle point $\bx^{**}(\tau)$ of this functional.
%We notice that, if the underlying pseudodynamics used to sample the paths is given by a first order Langevin dynamics, this analogy between quantum statistic and classical statistical mechanics have been already worked out in the \emph{stochastic quantization} approach   by Parisi and Wu\cite{parisi1981perturbation} in the context of Quantum Field Theory.
%What is shown in Ref.~\onlinecite{isakov2015understanding} is that, for a double well model, we can identify the instanton $\bx^{**}(\tau)$ to be the transition state of the PIMC pseudodynamics.
%Indeed after the classical mapping the original quantum tunneling event has been recast into a thermally activated escape event.
%
%According to Kramers theory\cite{hanggi1990reaction} the escape rate is $k \propto e^{- F / k_B T}$, and therefore $k \propto \Delta^2$ if standard closed path integrals are used, whereas  $k \propto \Delta$ if the paths are opened.
%In the following, in short, we will address the first approach as simply PBC, while the latter as OBC.

Let us start with the PIMC ( or equivalently PIMD) simulation, where we samples paths $\bx(\tau,t)$ at each update along the simulation time axis $t$, and these paths are distributed according the  functional $ S(\bx(\tau)) $ as in Eq.~(\ref{e:pi}).
We notice that, if the underlying pseudodynamics used to sample the paths is given by a first order Langevin dynamics, 
$\partial \bx(\tau,t) / \partial t = - \delta S / \delta \bx(\tau,t) + \eta(\tau,t)$
this analogy between quantum statistic and classical statistical mechanics have been already worked out in the \emph{stochastic quantization} approach   by Parisi and Wu\cite{parisi1981perturbation} in the context of Quantum Field Theory.
Here, the velocity of the (deformations of) path $\partial \bx(\tau,t) / \partial t$ , are linked to the generalized forces $\delta S / \delta \bx(\tau,t)$ and a gaussian white noise $\eta(\tau,t)$ satisfying the obvious fluctuation-dissipation relation.
If the system displays bi-stable minima, then the transition state of the pseudodynamics is given by the point $\bx_{TS}(\tau)$ satisfying $\delta S( \bx_{TS}(\tau))/ \delta \bx(\tau) = 0$ and which is not already in one of the attraction basins corresponding to the two minima\cite{parisi1981perturbation,sega2007quantitative,autieri2009dominant,mazzola2011fluctuations}.

Finding this transition state is generally very complicated, but in the case of a double well potential $V(x)$ this can be done analytically.
Here, the dominant contribution to the integral comes from the stationary action path $\bx^{**}(\tau)$ (determined exactly by the condition
 $\delta S( \bx(\tau))/ \delta \bx(\tau) = 0$) which is called instanton\cite{coleman_fate_1977,forkel,Chudnovsky-book}.
 This trajectory in imaginary time corresponds to a particle moving in the inverted potential $-V(x)$ (see Fig.~\ref{fig:cart}) and it is possible to evaluate the action $S$ at this point.

Following Ref.~\onlinecite{isakov2015understanding} the amplitude is given by
\begin{equation}
  \exp({-S[\bx^*(\tau)]})  \propto \Delta \quad \textrm(instanton),
\end{equation} 
where $\bx^*(\tau)$ is  the open trajectory which connects the two classical turning points under the barrier, near the minima, and $\Delta$ is the \emph{tunneling} splitting.
%with $0 \le \tau < \beta = \hbar/k_B T$.
Notice that, when computing the (diagonal) density matrix $\rho(x)$ periodic boundary conditions (PBC) in imaginary time are required.
Now the integral over the \emph{closed} paths it is dominated by the imaginary time trajectory $\bx^{**}(\tau)$ that moves under the barrier starting, reaches the turning point, and returns.
Therefore the saddle point extimation of the integral  gives a {\it squared} tunneling amplitude
\begin{equation}
  \exp({-S[\bx^{**}(\tau)]}) \propto \Delta^2  \quad \textrm(double~ instanton),
\end{equation}
due to the cost of creating an instanton and an anti-instanton (see Fig.~\ref{fig:cart}).
Coming back to the PIMC pseudodynamics, 
according to Kramers theory\cite{hanggi1990reaction}, the escape rate is $k \propto e^{- S(\bx_{TS})}$, and therefore $k \propto \Delta^2$ if standard closed path integrals are used, whereas  $k \propto \Delta$ if the paths are opened.
In the following, in short, we will address the first approach as simply PBC, while the latter as OBC.

In this paper we extend the study of Ref.~\onlinecite{isakov2015understanding} from spin hamiltonians to continuous variables models, which are relevant for realistic quantum simulations.
We demonstrate that the same quadratic speedup, in sampling tunneling events, occurs in a double well model, in which we can tune separately the  width and the height of the energy barrier.
We also show that it is possible to sample from ground state distribution $|\psi_0(x)|^2$ by considering the center of the open-path $\bx^{*}(\tau =\beta/2)$, whereas the tails $\bx^{*}(\tau =0)$,  $\bx^{*}(\tau =\beta)$ sample from the ground state distribution $\psi_0(x)$.
Moreover, in the armonic case it is also possible to sample from correct finite temperature distribution $\rho_\beta(x)$ by using the center of the path.

%Notice that the idea of using open path integrals in the realm of realistic simulations is certainly not new.
%Indeed open paths have been employed at nearly zero temperature, i.e. long projecting imaginary time $\beta$, to find the ground state (this method is called \emph{path integral ground state})\cite{pigs}, or to compute off-diagonal operators, such as the momentum, in superfluid helium\cite{pollock} or liquid water\cite{morrone2007}.
%To this end they have also been used to study nuclear quantum effects in water in the pioneering work of Morrone and Car\cite{carwater}.
%
%Nevertheless the aim of this study is instead to demonstrate that open paths could also greatly reduce correlation times in path integral based simulations. 
%Therefore we propose that OBC should be used not only in the context of Simulated Quantum Annealing but also in realistic material simulations.
In the following we provide some prototypical examples to demonstrate this feature.

\section{Double well potential}
\begin{figure}[t]
 \includegraphics[width=1\columnwidth]{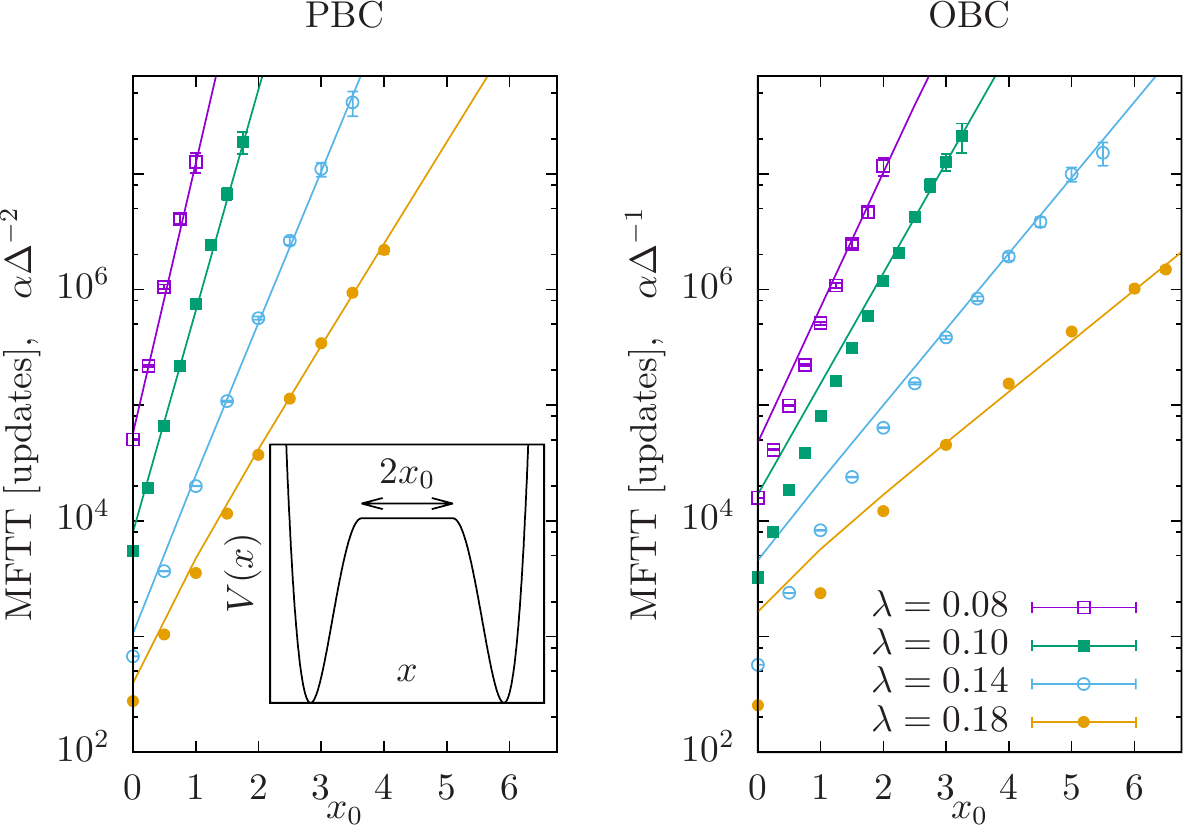}
 \caption{ (color online). Average  MFTT tunneling time with PIMC (for PBC and OBC) as a function of $x_0$ for different values of $\lambda$, at $\beta=20$, corresponding to a temperature always much lower than the barrier height. 
  The inset shows the shape of double well potential $V(x)$, which barrier width (at top) is $2 x_0$. 
   Notice that for OBC, the measured MFTT is smaller than the one predicted by the $1/\Delta$ formula, when the tunneling rate is large. This happens because both the 
 $\bx^{*}(\tau)$ and the $\bx^{**}(\tau)$ channel contribute to the tunneling.
 }
 \label{fig:gaps}
 \end{figure}

Let us consider the following one-dimensional double well potential,
\begin{align}
\label{e:pot}
V(x) =\begin{cases} 
      \lambda (x-x_0)^4 - (x-x_0)^2, & x\geq x_0 \\
      0, & -x_0\le x\le x_0 \\
       \lambda (x+x_0)^4 - (x+x_0)^2, & x\leq -x_0
   \end{cases}
\end{align}
with $\lambda,x_0 > 0$.
%This potential represents a generalization of Ref.~\onlinecite{isakov2015understanding} as now
 We can separately tune  the width and the height of the barrier, varying $\lambda$ and $x_0$. The energy barrier is $\Delta V = 1/4 \lambda$, and the distance between the two minima is 
$d  = 2 (x_0 + \sqrt{1/ 2\lambda})$ (see inset of Fig.~\ref{fig:gaps})
Decreasing $\lambda$ reduces the energy splitting $\Delta$, as the two wells become deeper and more separated.
The parameter $x_0$  only increases the well separation but doesn't change the potential energy barrier.
%%%Our Hamiltonian reads 
%\begin{equation}
%H = {d^2 \over dx^2} + V(x),
%\end{equation}
%which equals to se the mass to $1/2$ and $\hbar=1$.
 
 Following Ref.~\onlinecite{isakov2015understanding} we measure the mean first tunneling time (MFTT), defined as the number of MC updates required to find the system in the right well, if  the particle is localized in the left one at the beginning of the simulation.
 From Fig.~\ref{fig:gaps} we see that the MFTT scales as $1/\Delta^2$ when PBC are used, whereas it scales as  $1/\Delta$ for OBC, as the parameters $x_0$ and $\lambda$ change.
  The exact gap value are obtained using a discrete variable representation (DVR) technique\cite{colbert1992novel}.
  This scaling relation holds for PIMC with local Metropolis updates and PIMD with first and second order Langevin thermostats.
As far as standard PIMC is concerned, this means that the scaling of tunneling rate in a double well model $k \propto \Delta^2$ is correctly reproduced\cite{weiss1987incoherent}.
Therefore, we expect equilibrium PIMC or PIMD simulations to faithfully describe tunneling rate ratios as a function of the various control parameters, such as density, isotope masses and the accuracy of the potential energy surface $V(x)$ which can be obtained by different electronic techinques\cite{sprik1993,ceriotti2013nuclear,zen2015ab,del_ben_bulk_2013,kuhnejctc2009}.

We find that for sufficiently low temperatures, it is possible to sample from the correct ground state distribution $|\psi_0(x)|^2$ by considering the center of the open-path $\bx^{*}(\tau =\beta/2)$, whereas the tails $\bx^{*}(\tau =0)$,  $\bx^{*}(\tau =\beta)$ sample from the ground state distribution $\psi_0(x)$. 
Notice that, in the PIGS\cite{pigs} approach this would be the mixed distributution $\psi_0(x) \psi_T(x)$, but in this case the trial wavefunction is $\psi_T(x)=1$.
%This property follows directly from the PIGS formalism\cite{pigs}.
   \begin{figure}[t]
 \includegraphics[width=1\columnwidth]{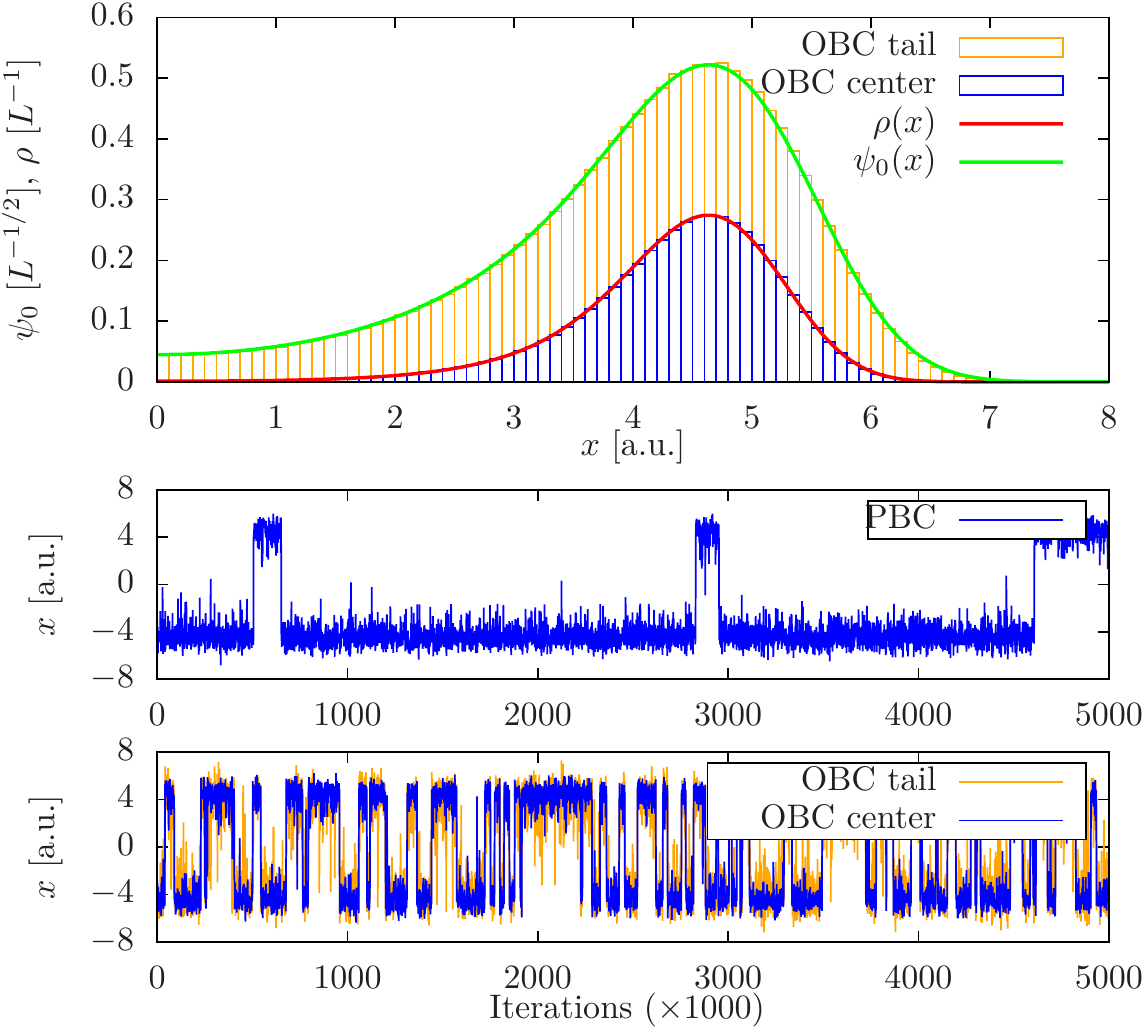}
 \caption{ (color online). Top panel: position distributions (histograms) obtained considering the center (blue) or the tail (orange) of the OBC path. The distributions are area-normalized respectively with the exact $\rho(x)\approx |\psi_0|^2$ distribution (red) and the exact ground state $\psi_0(x)$ (green). 
We plot only for $x>0$ and we use $x_0=3$ and $\lambda = 0.14$ in Eq.~(\ref{e:pot}). 
 The difference between the sampled distrubutions and the reference ones are negligible. We perform simulations at low temperatures, $\beta=20 \gg \Delta V$.
 Middle and lower panel: the position of the particle as the simulation progresses for PBC and OBC (both for center and tail). As expected the tunneling rate is much larger for OBC.
 }
 \label{fig:dw}
 \end{figure}
In Fig.~\ref{fig:dw} we see that it is possible to sample from the exact equilibrium distribution $\rho(x)\approx  |\psi_0|^2$ while having a considerable speed-up in the sampling, using OBC and considering the replicas located in the center of the path.

\section{Harmonic Oscillator and finite temperature simulations}

\begin{figure}[t]
 \includegraphics[width=1\columnwidth]{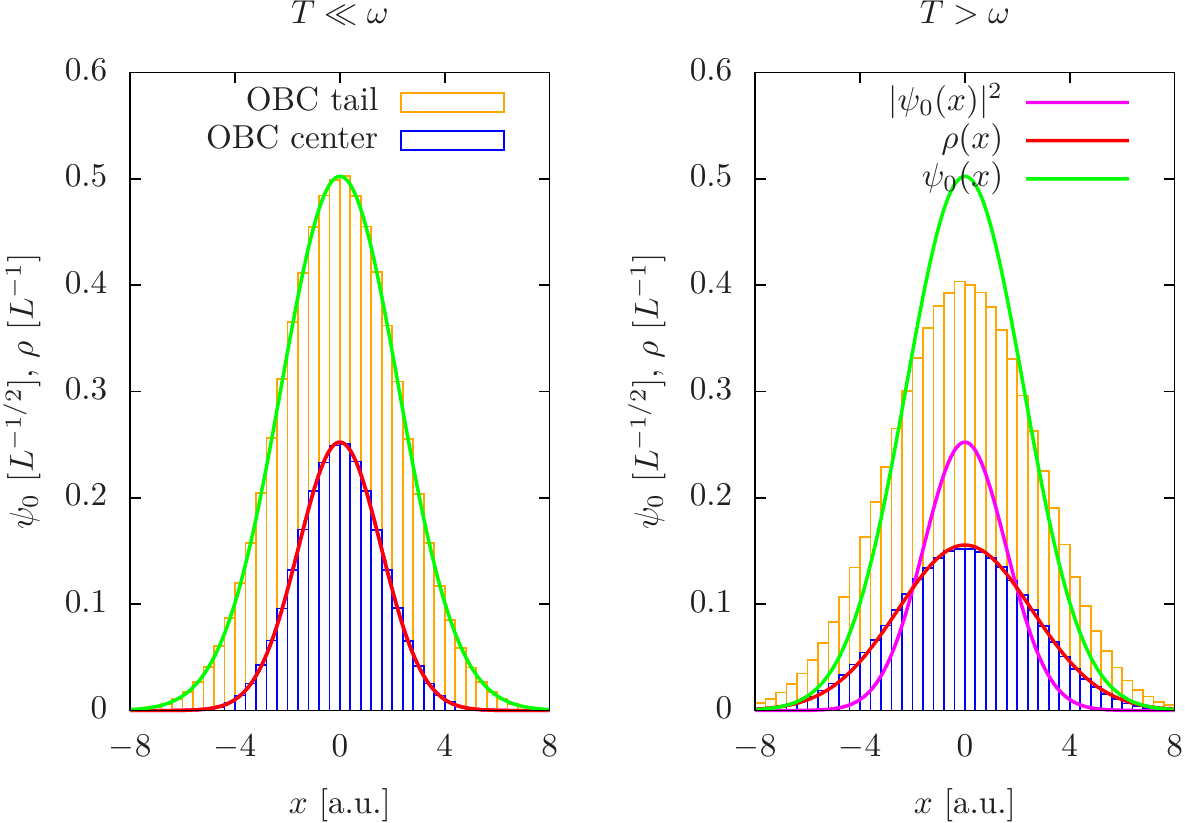}
 \caption{ (color online). Position distribution for harmonic potential at two different temperatures. We sample the distribution at the center (blue) and at the tails (orange) of the open path. At low temperature (left panel) the distributions coincide respectively with the exact $|\psi_0|^2$ (magenta) and $|\psi_0|$ (green) ones. Interestingly the center of the path still sample the correct finite temperature distribution $\rho(x)$ (red) at larger temperature $T>\omega$.
 }
 \label{fig:ho}
 \end{figure}

We next investigate the ability of OBC to simulate finite temperature properties.
We consider a harmonic potential of the form
%\begin{equation}
$V(x)= 1/2 ~m \omega^2 x^2$,
%\end{equation}
with $m=1/2$ and $\omega=0.4$.
We perform simulations at temperatures respectively smaller and larger than energy gap $\omega$.
From Fig.~\ref{fig:ho} we see that the center of the OBC path still samples the exact thermal distribution $\rho(x)$, which differs from the simple ground state density $\psi_0(x)^2$ at large temperature.
%We notice that now, at large temperature, the tails of the path do not correctly sample  the ground state distribution  $\psi_0(x)$.
Unfortunately this property does not hold in the general case, for example,  in the double well model considered above, we make an error of $\approx 10\%$ in the sampled distribution, at a large temperatures $T \sim \Delta V$.  Indeed, in this case, the trade-off between accuracy and speed-up has to be carefully checked for non-zero temperature simulations.

\section{Conclusions}

We have studied the autocorrelation properties of path integral based  equilibrium simulations with periodic and open boundary conditions in imaginary time.
While the former technique is widely used to simulate nuclear quantum effects at finite temperature, the latter is also a well established approach --- although less popular compared to its PBC counterpart --- used to calculate ground state properties.

We have numerically demonstrated that the autocorrelation time of open paths simulations can be much smaller than the corresponding periodic case.
In a double well model, characterized by quantum tunneling mechanism, we obtain a clear quadratic speedup as a function of the tunneling  energy splitting $\Delta$, which in turn is given by the shape of the potential energy barrier. This holds in both  continuos space and  spin models.
We also provide an intuitive explanation based on the role of \emph{instantons} in the PIMC pseudodynamics.

Therefore we propose that, in all cases when the ground state approximation to the finite temperature partition function holds, open path integral simulations should be used and will accelerate the sampling.
The computational gain of using open paths is clearly system dependent, but is expected to be particularly large when rare quantum tunneling events become important.

%To summarize, we propose that 
%\emph{open}-path integral simulations will greatly reduce autocorrelation times compared to standard path-integral simulation with ring-polymer (PBC in imaginary time), for simulating low-temperature NQE.
%In the case of simple double well model we obtain a clear quadratic speed-up as a function of the potential parameters which control the....

\section{Acknowledgements}
This work was supported by the European Research Council through ERC
Advanced Grant SIMCOFE by the Swiss National Science Foundation through
NCCR QSIT, and by Microsoft Research. This paper is based upon work
supported in part by ODNI, IARPA via MIT Lincoln Laboratory Air Force
Contract No. FA8721-05-C-0002. The views and conclusions contained
herein are those of the authors and should not be interpreted as necessarily
representing the official policies or endorsements, either expressed
or implied, of ODNI, IARPA, or the U.S. Government. The U.S. Government
is authorized to reproduce and distribute reprints for Governmental
purpose not-withstanding any copyright annotation thereon.
We acknowledges useful discussions with P. Faccioli and S.Sorella.
MT acknowledges hospitality of the Aspen Center for Physics, supported by NSF grant PHY-1066293.

%\bibliography{biball,bibQA,bibPIMD,Bibliography}

\begin{thebibliography}{10}

\bibitem{ceriotti2016nuclear}
M.~Ceriotti, W.~Fang, P.~G. Kusalik, R.~H. McKenzie, A.~Michaelides, M.~A.
  Morales, and T.~E. Markland,
\newblock Chemical reviews  (2016).

\bibitem{carwater}
J.~A. Morrone and R.~Car,
\newblock Phys. Rev. Lett. {\bf 101}, 017801 (2008).

\bibitem{ceriotti2012efficient}
M.~Ceriotti and D.~E. Manolopoulos,
\newblock Physical review letters {\bf 109}, 100604 (2012).

\bibitem{ceriotti2013nuclear}
M.~Ceriotti, J.~Cuny, M.~Parrinello, and D.~E. Manolopoulos,
\newblock Proceedings of the National Academy of Sciences {\bf 110}, 15591
  (2013).

\bibitem{benoit1998tunnelling}
M.~Benoit, D.~Marx, and M.~Parrinello,
\newblock Nature {\bf 392}, 258 (1998).

\bibitem{meng2015direct}
X.~Meng, J.~Guo, J.~Peng, J.~Chen, Z.~Wang, J.-R. Shi, X.-Z. Li, E.-G. Wang,
  and Y.~Jiang,
\newblock Nature Physics {\bf 11}, 235 (2015).

\bibitem{richardson2016concerted}
J.~O. Richardson, C.~P{\'e}rez, S.~Lobsiger, A.~A. Reid, B.~Temelso, G.~C.
  Shields, Z.~Kisiel, D.~J. Wales, B.~H. Pate, and S.~C. Althorpe,
\newblock Science {\bf 351}, 1310 (2016).

\bibitem{bonev2004quantum}
S.~A. Bonev, E.~Schwegler, T.~Ogitsu, and G.~Galli,
\newblock Nature {\bf 431}, 669 (2004).

\bibitem{pickard_structure_2007}
C.~J. Pickard and R.~J. Needs,
\newblock Nature Physics {\bf 3}, 473 (2007).

\bibitem{labet_fresh_2012}
V.~Labet, R.~Hoffmann, and N.~W. Ashcroft,
\newblock The Journal of Chemical Physics {\bf 136}, 074504 (2012).

\bibitem{morales_nuclear_2013}
M.~A. Morales, J.~M. {McMahon}, C.~Pierleoni, and D.~M. Ceperley,
\newblock Physical Review Letters {\bf 110}, 065702 (2013).

\bibitem{mazzola_unexpectedly_2014}
G.~Mazzola, S.~Yunoki, and S.~Sorella,
\newblock Nature Communications {\bf 5}, 3487 (2014).

\bibitem{eremets_conductive_2011}
M.~I. Eremets and I.~A. Troyan,
\newblock Nature Materials {\bf 10}, 927 (2011).

\bibitem{howie_mixed_2012}
R.~T. Howie, C.~L. Guillaume, T.~Scheler, A.~F. Goncharov, and E.~Gregoryanz,
\newblock Phys. Rev. Lett. {\bf 108}, 125501 (2012).

\bibitem{dalladay2016evidence}
P.~Dalladay-Simpson, R.~T. Howie, and E.~Gregoryanz,
\newblock Nature {\bf 529}, 63 (2016).

\bibitem{mazzola2015distinct}
G.~Mazzola and S.~Sorella,
\newblock Physical Review Letters {\bf 114}, 105701 (2015).

\bibitem{sorella2016accelerated}
S.~Sorella and G.~Mazzola,
\newblock arXiv preprint arXiv:1605.08423  (2016).

\bibitem{knudson2015direct}
M.~Knudson, M.~Desjarlais, A.~Becker, R.~Lemke, K.~Cochrane, M.~Savage,
  D.~Bliss, T.~Mattsson, and R.~Redmer,
\newblock Science {\bf 348}, 1455 (2015).

\bibitem{RevModPhys.67.279}
D.~M. Ceperley,
\newblock Rev. Mod. Phys. {\bf 67}, 279 (1995).

\bibitem{ceriotti_nuclear_2009}
M.~Ceriotti, G.~Bussi, and M.~Parrinello,
\newblock Physical Review Letters {\bf 103}, 030603 (2009).

\bibitem{ceriotti}
M.~Ceriotti, M.~Parrinello, T.~Markland, and D.~Manolopoulos,
\newblock Journal of Chemical Physics {\bf 133}, 124104 (2010).

\bibitem{markland2008efficient}
T.~E. Markland and D.~E. Manolopoulos,
\newblock The Journal of chemical physics {\bf 129}, 024105 (2008).

\bibitem{PhysRevE.93.043305}
C.~John, T.~Spura, S.~Habershon, and T.~D. K\"uhne,
\newblock Phys. Rev. E {\bf 93}, 043305 (2016).

\bibitem{kapil2016accurate}
V.~Kapil, J.~VandeVondele, and M.~Ceriotti,
\newblock The Journal of chemical physics {\bf 144}, 054111 (2016).

\bibitem{pigs}
A.~Sarsa, K.~E. Schmidt, and W.~R. Magro,
\newblock J. Chem. Phys. {\bf 113}, 1366 (2000).

\bibitem{doi:10.1021/jp4015178}
S.~Constable, M.~Schmidt, C.~Ing, T.~Zeng, and P.-N. Roy,
\newblock The Journal of Physical Chemistry A {\bf 117}, 7461 (2013),
\newblock PMID: 23738885.

\bibitem{pilati2012bosonic}
S.~Pilati and M.~Troyer,
\newblock Physical review letters {\bf 108}, 155301 (2012).

\bibitem{pollock}
D.~Ceperley and E.~Pollock,
\newblock Can. J. Physics {\bf 65}, 1416 (1987).

\bibitem{bertaina2016one}
G.~Bertaina, M.~Motta, M.~Rossi, E.~Vitali, and D.~Galli,
\newblock Physical review letters {\bf 116}, 135302 (2016).

\bibitem{roggero2013dynamical}
A.~Roggero, F.~Pederiva, and G.~Orlandini,
\newblock Physical Review B {\bf 88}, 094302 (2013).

\bibitem{cuervo2005path}
J.~E. Cuervo, P.-N. Roy, and M.~Boninsegni,
\newblock The Journal of chemical physics {\bf 122}, 114504 (2005).

\bibitem{morrone2007}
J.~A. Morrone, V.~Srinivasan, D.~Sebastiani, and R.~Car,
\newblock The Journal of Chemical Physics {\bf 126} (2007).

\bibitem{PhysRevLett.82.4745}
S.~Baroni and S.~Moroni,
\newblock Phys. Rev. Lett. {\bf 82}, 4745 (1999).

\bibitem{carleo2010reptation}
G.~Carleo, F.~Becca, S.~Moroni, and S.~Baroni,
\newblock Physical Review E {\bf 82}, 046710 (2010).

\bibitem{cmd}
J.~Cao and G.~A. Voth,
\newblock The Journal of Chemical Physics {\bf 99} (1993).

\bibitem{rpmd}
I.~R. Craig and D.~E. Manolopoulos,
\newblock The Journal of Chemical Physics {\bf 121} (2004).

\bibitem{voth}
S.~Jang, A.~V. Sinitskiy, and G.~A. Voth,
\newblock The Journal of Chemical Physics {\bf 140} (2014).

\bibitem{shorttime}
B.~J. Braams and D.~E. Manolopoulos,
\newblock The Journal of Chemical Physics {\bf 125} (2006).

\bibitem{hele}
T.~J.~H. Hele, M.~J. Willatt, A.~Muolo, and S.~C. Althorpe,
\newblock The Journal of Chemical Physics {\bf 142} (2015).

\bibitem{farhi2000quantum}
E.~Farhi, J.~Goldstone, S.~Gutmann, and M.~Sipser,
\newblock arXiv preprint quant-ph/0001106  (2000).

\bibitem{Farhi20042001}
E.~Farhi, J.~Goldstone, S.~Gutmann, J.~Lapan, A.~Lundgren, and D.~Preda,
\newblock Science {\bf 292}, 472 (2001).

\bibitem{johnson2011quantum}
M.~Johnson, M.~Amin, S.~Gildert, T.~Lanting, F.~Hamze, N.~Dickson, R.~Harris,
  A.~Berkley, J.~Johansson, P.~Bunyk, et~al.,
\newblock Nature {\bf 473}, 194 (2011).

\bibitem{boixo2014}
S.~Boixo, T.~F. Ronnow, S.~V. Isakov, Z.~Wang, D.~Wecker, D.~A. Lidar, J.~M.
  Martinis, and M.~Troyer,
\newblock Nature Physics {\bf 10}, 218 (2014).

\bibitem{ronnow2014defining}
T.~F. R{\o}nnow, Z.~Wang, J.~Job, S.~Boixo, S.~V. Isakov, D.~Wecker, J.~M.
  Martinis, D.~A. Lidar, and M.~Troyer,
\newblock Science {\bf 345}, 420 (2014).

\bibitem{isakov2015understanding}
S.~V. Isakov, G.~Mazzola, V.~N. Smelyanskiy, Z.~Jiang, S.~Boixo, H.~Neven, and
  M.~Troyer,
\newblock arXiv preprint arXiv:1510.08057  (2015).

\bibitem{jiang2016scaling}
Z.~Jiang, V.~N. Smelyanskiy, S.~V. Isakov, S.~Boixo, G.~Mazzola, M.~Troyer, and
  H.~Neven,
\newblock arXiv preprint arXiv:1603.01293  (2016).

\bibitem{parisi1981perturbation}
G.~Parisi, Y.-s. Wu, et~al.,
\newblock Scientia Sinica {\bf 24}, 483 (1981).

\bibitem{sega2007quantitative}
M.~Sega, P.~Faccioli, F.~Pederiva, G.~Garberoglio, and H.~Orland,
\newblock Physical review letters {\bf 99}, 118102 (2007).

\bibitem{autieri2009dominant}
E.~Autieri, P.~Faccioli, M.~Sega, F.~Pederiva, and H.~Orland,
\newblock The Journal of chemical physics {\bf 130}, 064106 (2009).

\bibitem{mazzola2011fluctuations}
G.~Mazzola, S.~a~Beccara, P.~Faccioli, and H.~Orland,
\newblock The Journal of chemical physics {\bf 134}, 164109 (2011).

\bibitem{coleman_fate_1977}
S.~Coleman,
\newblock Phys. Rev. D {\bf 15}, 2929 (1977).

\bibitem{forkel}
H.~{Forkel},
\newblock ArXiv High Energy Physics - Phenomenology e-prints  (2000).

\bibitem{Chudnovsky-book}
E.~M. Chudnovsky and J.~Tejada,
\newblock {\em Macroscopic Quantum Tunneling of the Magnetic Moment},
\newblock Cambridge, UK: Cambridge University Press, 1998.

\bibitem{hanggi1990reaction}
P.~H{\"a}nggi, P.~Talkner, and M.~Borkovec,
\newblock Reviews of modern physics {\bf 62}, 251 (1990).

\bibitem{colbert1992novel}
D.~T. Colbert and W.~H. Miller,
\newblock The Journal of chemical physics {\bf 96}, 1982 (1992).

\bibitem{weiss1987incoherent}
U.~Weiss, H.~Grabert, P.~H{\"a}nggi, and P.~Riseborough,
\newblock Physical Review B {\bf 35}, 9535 (1987).

\bibitem{sprik1993}
K.~Laasonen, M.~Sprik, M.~Parrinello, and R.~Car,
\newblock The Journal of Chemical Physics {\bf 99}, 9080 (1993).

\bibitem{zen2015ab}
A.~Zen, Y.~Luo, G.~Mazzola, L.~Guidoni, and S.~Sorella,
\newblock The Journal of chemical physics {\bf 142}, 144111 (2015).

\bibitem{del_ben_bulk_2013}
M.~Del~Ben, M.~Sch\"{o}nherr, J.~Hutter, and J.~VandeVondele,
\newblock The Journal of Physical Chemistry Letters {\bf 4}, 3753 (2013).

\bibitem{kuhnejctc2009}
T.~D. K{\"u}hne, M.~Krack, and M.~Parrinello,
\newblock Journal of Chemical Theory and Computation {\bf 5}, 235 (2009).

\end{thebibliography}

%\begin{thebibliography}{99}

%\end{thebibliography}

\end{document}